\documentclass[aip,prl,superscriptaddress,amsmath,amssymb,preprint,graphicx,endfloats*]{revtex4}

\usepackage{graphicx}
\usepackage{dcolumn}
\usepackage{bm}
\usepackage{color}
\usepackage{amsmath}
\usepackage{float}
\usepackage{array}
\usepackage[section]{placeins}

\begin{document}

\title{Nanoscale wear and kinetic friction between atomically smooth surfaces sliding at high speeds}
\newcommand {\HGST}{HGST, a Western Digital Company, San Jose Research Center, San Jose, CA 95135 USA.}

\author{Sukumar Rajauria}
\altaffiliation{email: sukumar.rajauria@hgst.com}
\affiliation{\HGST}
\author{Sripathi V. Canchi}
\altaffiliation{email: sripathi.canchi@hgst.com}
\affiliation{\HGST}
\author{Erhard Schreck}
\affiliation{\HGST}
\author{Bruno Marchon}
\affiliation{\HGST}

\begin{abstract}
The kinetic friction and wear at high sliding speeds is investigated using the head-disk interface of hard disk drives, wherein, the  head and the disk are less than 10 $nm$ apart and move at sliding speeds of 5-10 $m/s$ relative to each other. While the spacing between the sliding surfaces is of the same order of magnitude as various AFM based fundamental studies on friction, the sliding speed is nearly six orders of magnitude larger, allowing a unique set-up for a systematic study of nanoscale wear at high sliding speeds. In a hard disk drive, the physical contact between the head and the disk leads to friction, wear and degradation of the head overcoat material (typically diamond like carbon). In this work, strain gauge based friction measurements are performed; the friction coefficient as well as the adhering shear strength at the head-disk interface are extracted; and an experimental set-up for studying friction between high speed sliding surfaces is exemplified.
\end{abstract}

\date{\today}
\maketitle



\maketitle
Tribology at nano and micro scales is a field of growing interest. An in-depth understanding of friction and its properties is essential for the design of miniaturized devices (micro-electro mechanical systems) \cite{Dowson}. Phenomenological laws of dry friction were established by Amontons and Coulomb, around 200 years ago, and it was empirically determined that the frictional force between two macroscopic bodies is linearly proportional to the applied load, and is independent of the macroscopic `apparent' contact area \cite{Amontons1699}. Later, Bowden and Tabor proposed that at the microscopic level, the friction force arises due to the shear strength of the contacting junction, which in turn is proportional to the `real' contact area, and verified it through various experimental and theoretical studies \cite{Bowden,HomolaWear90,BermanTL98,MoNature,CarpickJCIS99}. For adhering surfaces the friction force is a sum of both the load dependent and real contact area dependent forces \cite{BermanTL98}. However, most of these frictional studies at nanometer scale are limited to sliding speeds of less than few tens of micrometers per seconds \cite{GneccoPRL00, Zworner,Bhushan}.

Understanding friction at high sliding speeds, and at the nanometer scale has been particularly challenging \cite{MordukhovichTL11, IsraelachviliRPP10}. It is desirable to have surfaces with sub-nanometer roughness that sustain minimal wear during high speed contact. The head-disk interface of hard disk drives provides a unique platform for such studies.  In a hard disk drive, every intentional or unintentional contact between the head and the disk leads to friction and physical wear of the head overcoat layer, which typically is a diamond like carbon \cite{DaiIEEE03}. The head has a local microscale heater which sets both the sub-nanometer vertical clearance and a few tens of square microns of `apparent' contact area with the disk, which is sliding at a speed of 5-10 m/s relative to the head. These sliding speeds are six orders of magnitude higher than AFM based studies \cite{SuhTL06,TangJAP00}. While the current experimental results are motivated and illustrated using the head disk interface and the associated drive components, this unique setup may be extended to perform fundamental studies on nanoscale friction at high sliding speeds using other material interfaces.

\begin{figure}[h!]
\begin{center}
\includegraphics[width=0.5\textwidth,scale=0.5]{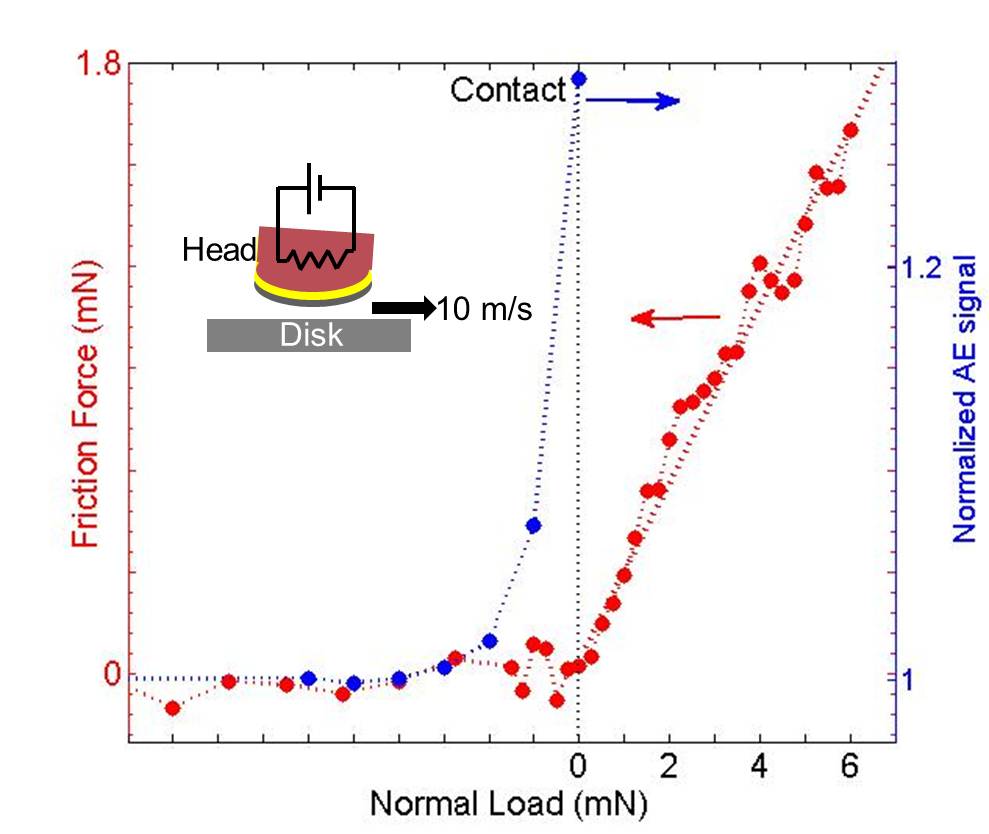}
\end{center}
\caption {Main: Normalized acoustic emission signal and the friction force between the head and the disk. Inset shows the schematic of high sliding speed head-disk interface. Vertical clearance and the normal load between the head and the disk is set using the microscale heater embedded inside the head.}
    \label{fig:1}
\end{figure}
\begin{figure*}[htbp]
\begin{center}
\includegraphics[width=1\textwidth]{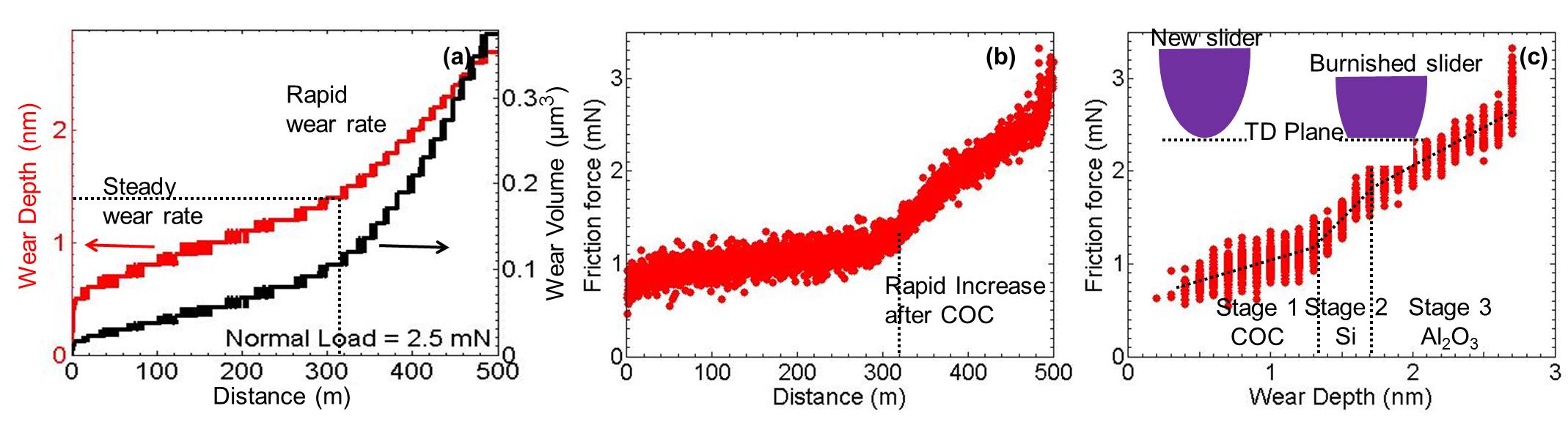}
\end{center}
\caption {(a) Wear depth and wear volume of the head overcoat as a function of distance traveled in contact at a normal load of 2.5 mN. (b) Friction force between the head and the disk as a function of distance traveled. (c) Friction force as a function of head wear depth.}
    \label{fig:2}
\end{figure*}

In this letter, a detailed experimental study of friction between the head and the disk and the resulting head overcoat wear is presented. The disk is fabricated by depositing a magnetic multilayer structure onto a glass substrate, and coated with $\approx$ 3 $nm$ amorphous nitrogenated carbon (protective overcoat layer), and finally covered with a molecular layer of perfluoropolyether polymer lubricant ($\approx$ 1 $nm$ thick). The head is coated with 1.4
$nm$ of diamond like carbon on top of a 0.3 $nm$ silicon based adhesive layer on an alumina coated substrate. The typical roughness $R_a$ of the disk and the head surfaces are both 0.4 $nm$. Thus, the head disk interface is a smooth carbon-lubricant-carbon high speed sliding interface.  The head surface is carefully etched such that an air lift force keeps it passively afloat at a fixed clearance over the disk \cite{JuangIEEE06}. The initial nominal  clearance (physical gap) between the head and the disk is typically 10 $nm$. Further, clearance is controlled using an embedded micro-heater in the head (inset Figure~\ref{fig:1}). The micro-heater generates a localized protrusion on the head surface, thus bringing it in contact with the disk. Contact between the head and the disk is detected using an acoustic emission (AE) sensor, which detects elastic propagating waves generated during the head-disk contact events \cite{CanchiAT12}. A calibrated strain gauge is instrumented to measure the friction force at the head disk interface along the sliding direction.

Figure~\ref{fig:1} shows a typical contact and friction measurement. As the micro-heater power increases (abscissa), a protrusion develops on the head (lateral dimensions $\approx$ 5 $\mu m$ $\times$ 2 $\mu m$ and radius of curvature of around 20 $mm$), bringing it closer to the disk. The contact is detected as a sharp increase in the AE signal. Further increase in the micro-heater power increases the applied normal load at the interface. Finite element based protrusion simulation is used to calculate the normal load $L$ at the interface \cite{ZhengTL10}. Simulations agree with the experiments within to 20$\%$ for various measurable quantities like initial flying height and protrusion geometry, and estimates that the normal load changes by 0.25 $mN$ per milliwatt of heater power beyond contact. Subsequent to contact, the data shows a linear increase in the friction force with increasing normal load at the interface, the slope being the friction coefficient. 

In most AFM based friction and wear studies, the wear is estimated ex-situ using either the AFM or a electron microscopy of the worn track, which limits the measurement of the wear depth/volume versus sliding distance to a few data points \cite{ChungTL03,KhurshudovTL96,BhushanAPL07}. Ideally, to gain better insight into the wear process an in-situ monitoring of the wear is desired. In this aspect, the head-disk interface has a unique feature with the micro-heater power calibrated precisely to measure the wear depth/volume in a continous manner during the experiment.  Figure~\ref{fig:2}a shows both the wear depth and  wear volume on the head overcoat as a function of the sliding distance traveled in contact with the disk at a constant normal load of 2.5 $mN$. The head overcoat wears rapidly by 0.5 $nm$ during the first 10 $m$ of traveled distance. Subsequently, the wear occurs at a steady rate of 0.0028 $nm/m$. The initial rapid head wear may be attributed to the rapid burnishing of outlying asperities on the head, as well as to the $sp^{2}$ enriched content on topical overcoat layer. After remaining in contact with disk for 300 $m$, the protective carbon overcoat is worn out completely, and the wear rate increases significantly. This result demonstrates the superior wear resistance of the carbon overcoat, and the protection it offers to the critical structures underneath. Figure~\ref{fig:2}b shows the friction force between the head and the disk for the experiment presented in Figure~\ref{fig:2}a. It shows that the friction force increases gradually with travelled distance, and rises sharply once the carbon overcoat is worn out. Figure~\ref{fig:2}c shows the friction force as a function of the overcoat wear depth. Based on the slopes, the wear is divided into three distinct `stages': stage-1 for the first 1.4 $nm$ (matching the thickness of the carbon overcoat); stage-2 extends for the next 0.3 $nm$ (matching the thickness of the Silicon based adhesive thin-film), and stage-3 for the wear of the alumina substrate material. It is interesting to note that the three stages, demarcated visually on the figure based on a change in the friction force, correspond excellently to the actual thickness of the carbon overcoat and the underlayer materials. It is noteworthy that stage-1 has the least slope, which qualitatively exemplifies that the wear resistance properties of the carbon overcoat are superior to those of the underlying materials. 

\begin{figure}[htbp]
\begin{center}
\includegraphics[width=3.5 in,height=1.9 in]{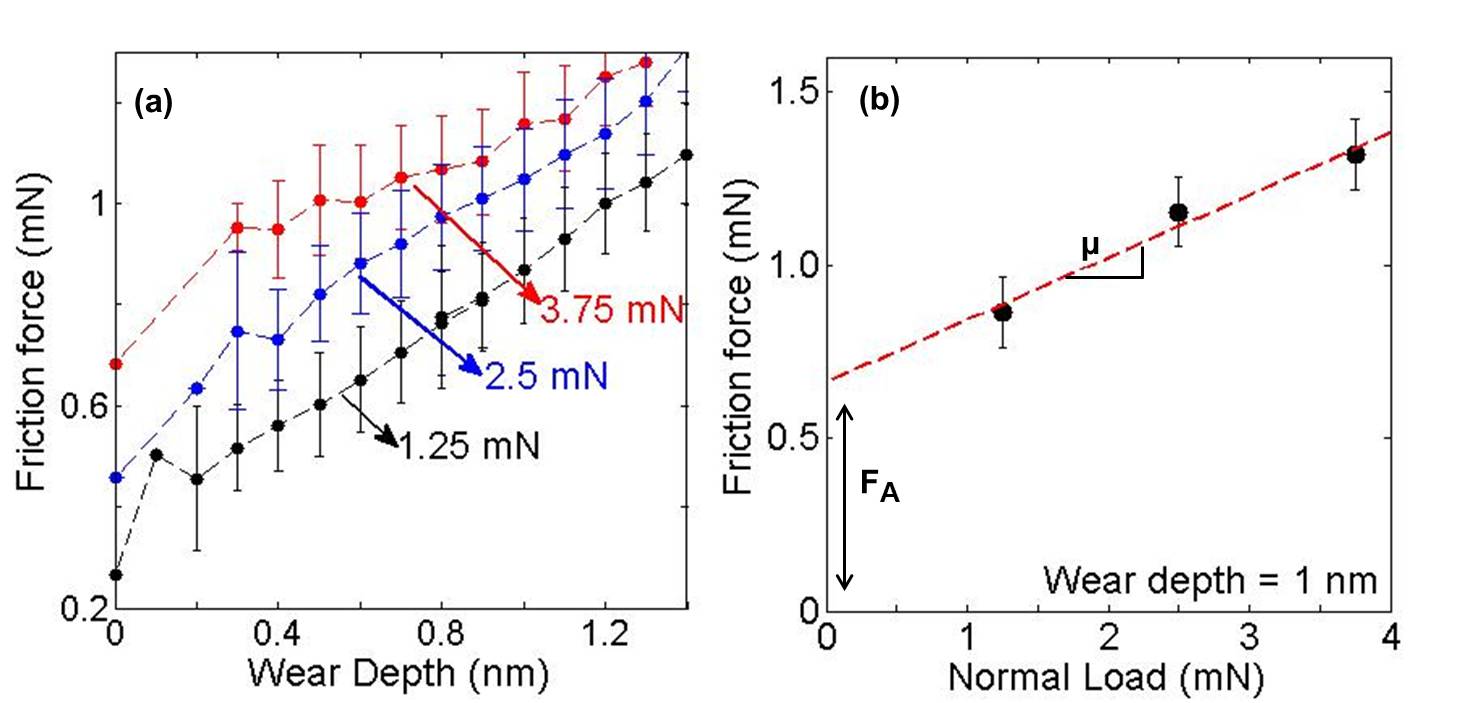}
\end{center}
\caption {(a) Friction force between the head and the disk at different normal load condition of 1.25, 2.5 and 3.75 mN respectively. (b) Friction force as a function of normal load, at a contant wear depth of 1 nm.}
    \label{fig:3}
\end{figure}

Figure~\ref{fig:3}a shows the friction force as a function of the overcoat wear on three different heads, where different normal load conditions are used for each case. For the range of normal load, the frictional heat is expected to increase the interface temperature by around 30-60 degrees Celsius. The experiment is restricted to stage-1, wherein only head carbon overcoat wear occurs. The friction force increases, both, with the overcoat wear depth and with increasing applied normal load condition (of 1.25, 2.5 and 3.75 $mN$ presented here). To further analyze the data, Figure~\ref{fig:3}b shows the friction force as a function of normal load, at a 
fixed wear depth of 1 $nm$ as obtained from Figure~\ref{fig:3}a, and the linear fit is also shown.  The aforementioned experimental data formalizes into the form, $F = \mu L + F_A$  \cite{BermanTL98}, where $F$ is the total friction at the interface, $\mu$ the coefficient of friction, $L$ the applied normal load, and $F_{A}$ the adhering component of friction. The slope of the linear fit in Figure~\ref{fig:3}a gives the coefficient of friction, and the ordinate intercept corresponds to the adhering friction at 
`no/negligible' applied normal load. This component of friction at `no/negligible' applied normal load may be attributed to the adhesive interaction between the atomically smooth interacting surfaces and is expected to scale with the real area of contact.

\begin{figure}[htbp]
\begin{center}
\includegraphics[width=3.5 in,height=1.9 in]{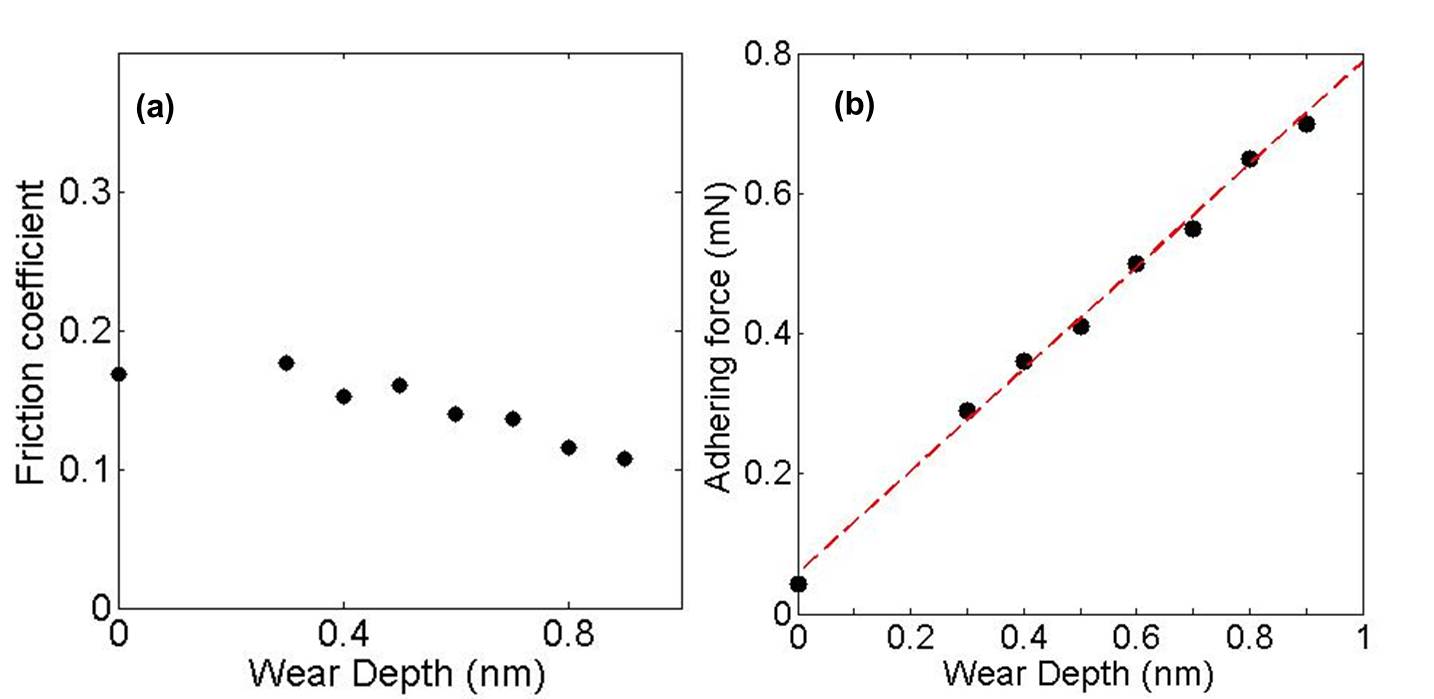}
\end{center}
\caption {(a) Shows the friction coefficient at different wear depth condition. (b) shows the adhering component of the friction force (F$_A$) under no load condition (L=0) as a function of wear depth. Red dashed line shows the linear fit to the data.}
    \label{fig:4}
\end{figure}

Figure~\ref{fig:4}a shows the friction coefficient (similarly as in Figure~\ref{fig:3}b) at different wear depth conditions, and it shows that the coefficient of friction changes little with increasing head wear. The friction coefficient obtained for this high sliding speed interface is similar to that reported in literature (ranging between 0.1-0.3), which were measured using an AFM for about six orders of magnitude slower sliding speed \cite{HaywardWear92, KonicekPRB12}.

Figure~\ref{fig:4}b shows the adhering component of the friction force $F_A$ as a function of the wear depth.  In our micro-heater geometry the real area of contact increases linearly with the overcoat wear depth. This is confirmed by a linear increase in the electric current between the head and disk as a function of the wear depth (see supplementary material) for a similar head-disk interface \cite{supplementary}. The data supports an excellent linear behavior between the adhering friction force and the wear depth. The non-zero offset at a wear depth of 0 $nm$ is due to the already existing contact area between the head and disk under pristine condition. 

This result exemplifies that the adhering component of frictional force $F_A$ takes the form, $F_A = \sigma A_0$, where $\sigma$ is the shear strength of the interface, and $A_0$ the real area of contact at `no/negligible' applied normal load conditions. Consequently, the slope of linear fit in Figure~\ref{fig:4} gives the shear strength $\sigma$ of the head disk interface. 
Estimating $A_0$ for a high sliding speed interface, which is  wearing with the sliding distance is non-trivial. Therefore, the shear strength is estimated from the measured apparent area of contact. The apparent area of contact is estimated by observing the burnish pattern on the head using an AFM. It is measured to be around $\approx$  18 $\mu m$ $\times$ 5 $\mu m$, which gives a  shear strength of around 9 $MPa$, which is in the range of measured values reported in literature \cite{HomolaWear90, IsraelachvilScience88}. It is noted that the apparent area of contact estimated from AFM agrees with the simulation to within 20$\%$.

In summary, a systematic study of friction and wear is performed for a high sliding speed interface using the experimental apparatus of hard disk drives. It is shown that the frictional force follows the dry frictional model even at high sliding speeds. In accordance with this model, a method to extract the coefficient of friction as well as the shear strength of the interface is outlined. This information and methodology is expected to be of great importance in understanding and improving the wear properties of nanoscale devices, which experience high sliding speeds.

With particular focus on the head disk interface of hard disk drives, a thorough understanding of the head wear process impacts the understanding of drive reliability, and hence holds great economic value to the industry. The results presented in this work demonstrate the superior wear resistance properties of the carbon overcoat and its importance in protecting the critical head elements underneath. The results also show that the adhering component of the friction force is important, increases with head wear, and can be a significant contributor to friction during contact events. 



\begin{thebibliography}{1}



\bibitem{Dowson}  D. Dowsin,  {\textit{History of Tribology} (Longman, London, 1979).}

\bibitem{Amontons1699} Amontons, G. 
De la resistance causee dans les machniines.
{\textit{Mem. Acad. R. A.} 275-282 (1699).}

\bibitem{BermanTL98} Berman, A., Drummond, C. and Israelachvili, J.
Amontons' law at the molecular level.
{\textit{Tribology Letters} $\textbf{4}$, 95-101 (1998).}

\bibitem{HomolaWear90} Homola, A. M., Israelachvili, J. N., McGuiggan P. M. and Gee M. L.
Fundamental experimental studies in tribology: The transition from “interfacial” friction of undamaged molecularly smooth surfaces to “normal” friction with wear.
{\textit{Wear} $\textbf{136}$, 65–83 (1990).}

\bibitem{MoNature} Mo, Y. F., Turner, K. T. and Szlufarska, I.
Friction laws at the nanoscale.
{\textit{Nature} $\textbf{457}$, 1116-1119 (2009).}

\bibitem{Bowden}  Bowden, F.P. and Tabor D.  {\textit{The friction and Lubrication of Solids} (Claredon, 1950).}

\bibitem{CarpickJCIS99} Carpick, R. W., Ogletree, D. F. and Salmeron, M.
A general equation for fitting contact area and friction vs load measurements.
{\textit{Journal of Colloid and Interface Science} $\textbf{211}$, 395-400 (1999).}


\bibitem{Zworner} Zworner, O., Holscher, H., Schwarz, U. D. and Wiesendanger, R.
The velocity dependence of frictional forces in point-contact friction.
{\textit{Applied Physics A-Materials Science and Processing } $\textbf{66}$, S263-S267 (1998).}

\bibitem{GneccoPRL00} Gnecco, E., Bennewitz, R., Gyalog, T., Loppacher, C., Bammerlin, M., Meyer, E. and Guntherodt, H. J. 
Velocity dependence of atomic friction.
{\textit{Physical Reviw Letters} $\textbf{84}$, 1172-1175 (2000).}

\bibitem{Bhushan} Bhushan, B.
Nano- to microscale wear and mechanical characterization using scanning probe microscopy.
{\textit{Wear} $\textbf{251}$, 1105-1123 (2001).}

\bibitem{IsraelachviliRPP10} Israelachvili, J., Min, Y., Akbulut, M.,  Alig, A., Carver, G., Greene, W., Kristiansen, K., Meyer, E., Pesika, N., Rosenberg, K. and Zeng, H.
Recent advances in the surface forces apparatus (SFA) technique.
{\textit{Rep. Prog. Phys.} $\textbf{73}$, 036601 (2010).}

\bibitem{MordukhovichTL11}   Lowrey, D. D., Tasaka, K., Kindt, J. H., Banquy, X., Belman, N., Min, Y., Pesika, N. S., Mordukhovich, G. and Israelachvili, J. N.  
High-Speed Friction Measurements Using a Modified Surface Forces Apparatus.
{\textit{Tribology Letters} $\textbf{42}$, 117-127 (2011).}


\bibitem{SuhTL06} Suh, A. Y., Mate, C. M., Payne, R. N. and Polycarpou, A. A.
Experimental and theoretical evaluation of friction at contacting magnetic storage slider-disk interfaces.
{\textit{Tribology Letters} $\textbf{23}$, 177-190 (2006).}

\bibitem{TangJAP00} Tang, H., Wang, L. P., Gui, J. and Kuo, D.
A study of dynamic friction at the head–disk interface.
{\textit{Journal of Applied Physics} $\textbf{87}$, 6152-6154 (2000).}

\bibitem{DaiIEEE03} Dai, Q., Yen, B. K., White, R. L., Peterson, P. J. and Marchon, B.
Toward an understanding of overcoat corrosion protection.
{\textit{IEEE Transactions on Magnetics} $\textbf{39}$, 2450-2452 (2003).}

\bibitem{JuangIEEE06} Juang, J. Y., Chen, D. and Bogy, D. B.
Alternate air bearing slider designs for areal density of 1 Tb$/$in$^{2}$.
{\textit{IEEE Transactions on Magnetics} $\textbf{42}$, 241-246 (2006).}

\bibitem{CanchiAT12} Canchi, S. V., Bogy, D. V., Wang, R. H. and Murthy A. N.
Parametric Investigations at the Head-Disk Interface of Thermal Fly-Height Control Sliders in Contact.
{\textit{Advances in Tribology} $\textbf{2012}$, 303071 (2012).}

\bibitem{ZhengTL10} Zheng, J. and Bogy, D.B.
Investigation of Flying-Height Stability of Thermal Fly-Height Control Sliders in Lubricant or Solid Contact with Roughness.
{\textit{Tribology Letters} $\textbf{38}$, 283-289  (2010).}

\bibitem{ChungTL03}  Chung, K. H. and Kim, D. E.   
Fundamental Investigation of Micro Wear Rate Using an Atomic Force Microscope.
{\textit{Tribology Letters} $\textbf{15}$, 135-144 (2003).}

\bibitem{KhurshudovTL96}  Khurshudov, A. G., Kato, K. and Koide H. 
Nano-wear of the diamond AFM probing tip under scratching of silicon, studied by AFM.
{\textit{Tribology Letters} $\textbf{2}$, 345-354 (1996).}

\bibitem{BhushanAPL07}  Bhushan, B. and Kwak K. J.
Velocity dependence of nanoscale wear in atomic force microscopy.
{\textit{Applied Physics Letters} $\textbf{91}$, 163113 (2007).}

\bibitem{HaywardWear92} Hayward, I.P. d, Singer, I.L. and Seitzma L.E.
Effect of roughness on the friction of diamond on cvd diamond coatings.
{\textit{Wear} $\textbf{157}$, 215 (1992).}

\bibitem{KonicekPRB12}  Konicek, A. R.,  Grierson, D. S., Sumant, A. V., Friedmann, T. A., Sullivan, J. P., Gilbert, P. U. P. A., Sawyer, W. G. and Carpick, R. W. 
Influence of surface passivation on the friction and wear behavior of ultrananocrystalline diamond and tetrahedral amorphous carbon thin films.
{\textit{Phys. Rev. B.} $\textbf{85}$, 155448 (2012).}


\bibitem{supplementary}   
See supplementary material at http://scitation.aip.org/content/aip/journal/apl for the overcoat wear as a fuction of current on a similar head-disk interface.

\bibitem{IsraelachvilScience88} Israelachvili, J.N. , McGuiggan, P.M. and Homola, A.M.  
Dynamic properties of molecularly thin liquid films.
{\textit{Science} $\textbf{240}$, 189-191 (1988).}

\end{thebibliography}
 \end{document}